\documentclass[article, twocolumn, notitlepage,superscriptaddress, longbibliography]{revtex4-2}
\usepackage{natbib}
\usepackage{graphicx}
\usepackage{amssymb}
\usepackage{amsmath}
\usepackage{ifthen}
\usepackage{braket}
\usepackage{xcolor}
\usepackage{bm}
\usepackage{bbm}
\usepackage{bbm}
\usepackage{maybemath}
\usepackage{comment}
\usepackage{siunitx}
\usepackage{float}
\usepackage{tabularray}
\usepackage{dcolumn}
\newcolumntype{d}[1]{D{.}{.}{#1}}
\usepackage{subfigure}

\definecolor{garrosgreen}{rgb}{0.1, 0.4, 0.1}
\definecolor{dartmouthgreen}{rgb}{0.05, 0.5, 0.06}
\definecolor{duelferred}{rgb}{0.7, 0.2, 0.1}
\definecolor{cambridgeblue}{rgb}{0.1, 0.3, 1.0}
\definecolor{oxfordblue}{rgb}{0.05, 0.2, 0.7}

\newcommand{\addrA}{Department of Chemistry, Physics and Materials Science, Fayetteville State University, Fayetteville, NC 28301, USA}

\begin{document}

\title{Effective Note-taking and its Impact on Learning Undergraduate Introductory Physics Courses}

\author{Chandra M. Adhikari} 
\email[ Corresponding author:\;]{cadhikari@uncfsu.edu}
\affiliation{\addrA}

\begin{abstract}
Taking notes during lectures is one of the required skills, among many others, that students need \textit{(i)} to master the topic covered in the lecture, \textit{(ii)} to actively engage in the learning process with minimal to no distractions, \textit{(iii)} to retain learned knowledge and skills for a longer time, and \textit{(iv)} in securing higher letter grades. To learn the role of notetaking in learning undergraduate-level introductory physics courses, we present a comparative study of students' achievement in a mid-terminal exam at a historically black Fayetteville State University (FSU)  \textit{(i)} when students were taught effective notetaking strategies, motivated them to prepare notes and let them use their self-prepared notes in a terminal exam versus \textit{(ii)} no notetaking scheme was implemented, keeping all other conditions the same. The no-notetaking scheme was used in 4 different sections over a few semesters, and the notetaking scheme was used at the same level of an introductory physics course in 3 different sections in other semesters. Students' scores in one of the mid-term exams are taken as measurement tools. Grade analysis indicates that effective notetaking enhances students' letter grades and lowers failure rates.
\end{abstract}

\maketitle


\section{\label{sec:level1} Introduction}

Based on the three-stage model of memory for information flow~\cite{Atkinson_SciAm_1971}, namely, \textit{(i)} sensory registers via visual, auditory, haptic, and so on  \textit{(ii)} short-term storage of information from temporary working memory and \textit{(iii)} long-term storage of information, learning can be perceived as encoding information permanently in the long-term memory~\cite {Balduccini_book_2011}. Sensory organs first perceive any information obtained from their source and then pass it to sensory memory. The information in sensory memory is transferred to working memory, also called short-term memory, and remains only temporarily or lost in no time. However, various control processes such as sufficient rehearsal, coding, decision-making, information retrieval, and retrieval strategies help to encode information in long-term storage and retain it there, which is otherwise lost or forgotten~\cite{Atkinson_SciAm_1971}. Learning new concepts, generating new ideas, acquiring skills, then storing them in long-term memory, and, when needed, retrieving them effortlessly and applying them is the primary goal of learning. 

One of the bonafide meanings of learning is storing the received knowledge, concepts, and gained skills in the long-term memory, retrieving them effortlessly as needed, and applying them in relevant contexts. Senses first perceive knowledge, information, course contents, or any other learning materials, pass to the sensory memory, go to the working memory, and finally, it either gets lost or gets stored in the long-term memory based on the learner's attention, elaborated rehearsal, and repetition~\cite{Felder__book_2016, Chew_CanPsychol_2021}. 
An effective learning process requires good organization of learning materials and attentiveness while learning. Students cannot sustain long-term attention to learning materials if activities that allow them to actively engage and contribute are omitted.  In a typical lecture with no activities, students' attentiveness peaks (70\%) at around 10 minutes and declines after that, reaching a minimum (20\%) at around 20 minutes~\cite{Felder__book_2016, Chew_CanPsychol_2021}. To maintain students' attentiveness at the maximum possible level throughout the lecture, an instructor can incorporate various activities, visuals, simulations, and other engaging elements at regular intervals.  Active processing helps move information from working memory to long-term memory~\cite{Sweller_2016JJarMac}, and taking notes is one way to do this.  In this study, we explore how learning can be improved by encouraging effective notetaking and adding activities during lectures.

Effects of notetaking on learning a secondary language have been extensively studied, with the literature showing that it helps in learning the second language (see a recent review in Ref.~\cite{Jin_SLA_2024}). In learning the second language vocabulary, Jin and Webb reported that vocabulary learning gains in the range of 12.5\% to 23.0\%~\cite{Jin_Webb_2021}. Authors in Ref.~\cite{Lee_Thesis_1998} presented that notetaking has even better impacts on second language listening comprehension for high school students. 
Campana conducted a study ~\cite{Campana_Thesis_2009}  on the effectiveness of using guided notes in high school physics by dividing the class into two groups. Therein, the author observed that the sections that were completed and used guided notes performed better academically.

To the best of our knowledge, no study has examined the role of effective notetaking in learning algebra-based undergraduate physics courses, as we presented in this research work to compare our findings; however, some works in the related science, technology, engineering, and mathematics (STEM) fields are available.  
Anastasiou et al. studied the effectiveness of concept maps in science learning~\cite{AnastasiouEtal_EPR_2024}. Therein the authors used data from 55 studies published between 1980 and 2020, including peer-reviewed journals and dissertations which covered various teaching settings, conditions, and methods and divulged that the concept maps were associated with increased science learning, but there was a significant heterogeneity in most subsets of data. 
In one of the studies made by Shi et al on 42 college students’ learning achievement and cognitive load in a 6-week lecture-based computer network course, the authors found that note taking helped in learning~\cite{Shi_Etal_AJET_2022}. In this regard, the authors demonstrated that among the three different notetaking styles observed, namely collaborative notetaking, laptop notetaking, and traditional longhand notetaking, collaborative notetaking was found to improve learning achievement. Recently, Fanguy studied the effects of collaboratively generated notes by students versus instructor-provided notes on recalling and academic writing performance and found that  student generated collaborative notes by students were more effective inenhancing academic writing, as well as in applying and developing related skills~\cite{Fanguy_AJET_2025}.

With the current technological advancement in the electronic industry, students are likely to suffer from many digital distractions in the learning environment, which interfere badly with learning~\cite{Weinstein_Book_2018}. In search of a better solution to minimize distraction in learning and increase attentiveness, we here present a case study on the role of effective notetaking in learning introductory undergraduate physics courses.  

\section{\label{sec:level2}   Effective Note-taking} 

 Notetaking significantly enhances deeper understanding, leading to metacognition, as writing notes helps the brain actively process and encode information, link ideas, integrate them with new knowledge, strengthen memory pathways, provide opportunities for reviewing, self-testing, and encourage reflection~\cite{Voyer_2020_CEP}. Notetaking offers students an elaborative rehearsal, supports analytical and critical thinking, and facilitates repetition and paraphrasing of the information, thereby enhancing their long-term memory~\cite{Bohay_2011_AJP}. Training students in notetaking skills, encouraging them to explore further, and helping them become effective notetakers are crucial. Taking an example of an undergraduate algebra-based physics course, the survey was conducted based on the hypothesis that notetaking helps in the learning process.

Notetaking is a crucial aspect of effective learning as it forces learners to engage with the learning material actively. It helps reinforce understanding and retention, encode information into long-term memory, provide opportunities for repetition and rehearsal, and organize information in a structured manner, making it easier to follow the logical flow of concepts.
The preferred way of learning or approaching a task varies from person to person based on how the mind receives and processes information from its senses. Learners' productivity, creativity, problem-solving skills, decision-making capability, quality of learning, and achievement can be increased and speed up by properly recognizing and understanding the learners’ preferred learning style~\cite{JaleelThomas2024}. Notes can be personalized to fit an individual's learning styles, integrating mnemonics, incorporating annotations and highlights that give the full picture, making a good sense of the concepts to the note-taker~\cite{Bain_Book_2023}. Well-prepared notes provide opportunities for review, revision, and repetition, essential for memorizing formulas, principles, and problem-solving techniques, improving memory~\cite{Salame_IJI_2024}.  

\begin{figure*}[tbh!]
\centering
\includegraphics[width=16.5 cm]{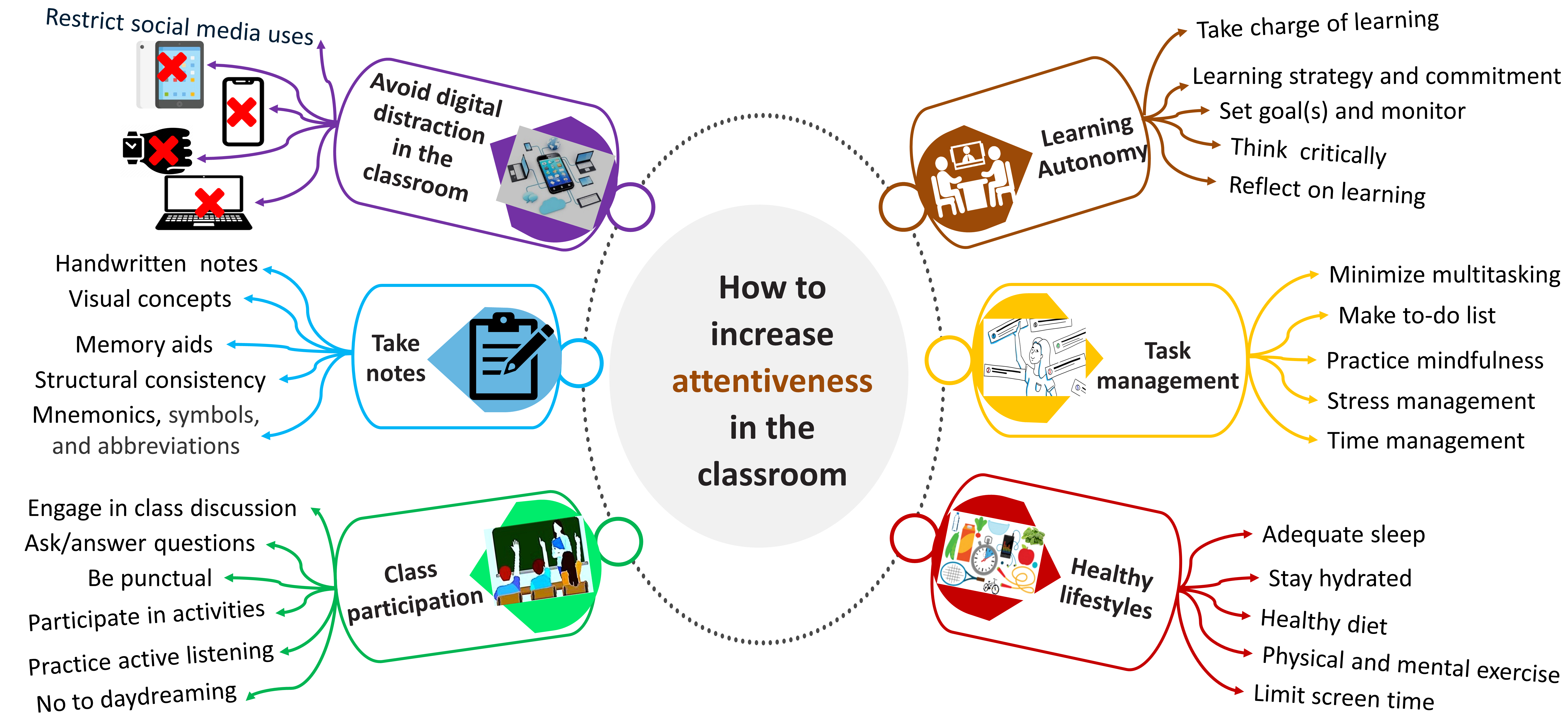}
\caption{The figure is an illustration of the mind-mapping method, showing ways to increase attentiveness in the classroom to improve the learning experience for students. \label{fig1}}
\end{figure*}  

Numerous notetaking methods are in practice to make the learning process easy, effective, and time-saving allowing students to choose the one that works best for them. Different notetaking methods, their effectiveness, and taking the best notes have been the center of research for quite a long time. For example, see Refs.[19-32]~\nocite{Stutts_2013, Rahmani_2011, Gonzalez_2018, Georgakis_2017, Wu_2018, Mueller_2018, Jones_2015, Boye_2012, Artz_JEC_2020, Dynarski_2017, Cohen_2013, Kauffman_2011, Reynolds_2016, Robin_1977}.
  
A brief review of different notetaking methods is in order. Cornell notetaking is one of the most commonly used techniques of taking notes, in which each page is divided into four segments, namely, a line at the top for the topic, a few lines at the bottom for the summary, and two columns divided for the middle part of the paper with a narrow left for key points and wider right column for an extensive detail of the key corresponding points~\cite{Pauk_2010}. Significantly structured notes with main topic(s) followed by indented subtopic(s), sub-subtopic(s), and detailed elaborations underneath in a logical manner forming a skeleton of the textbook chapter, called outline method, serves as an excellent study guide when preparing for tests ~\cite{Bui_JARMC_2015}. 
One can also use the boxing method of notetaking, which involves drawing boxes around related pieces of information to visually organize and categorize thoughts, concepts, or ideas of the same category by visually separating different topics under different boxes and titles, making it easier to review and memorize the notes for longer~\cite{Gupta_2024}.
The Zettelkasten method of notetaking is a personal knowledge management system, which serves well in creating an ever-growing web of knowledge over a certain topic, connecting fragmented note cards having one idea per note to a structured storing box, helping in the accumulation of knowledge on a specific topic over time~\cite{Kadavy_2021}.  In addition to being a memory retention tool, the Zettelkasten method fosters a deeper understanding of complex subjects and supports generating new ideas and insights, boosting creativity, productivity,  and adaptability in organizing and processing information.

 One of the most sophisticated, dynamic, and visually stimulating notetaking methods, in which a learner writes the main idea in the center of the page and branches out with connecting lines to subtopics, supporting details, and related concepts, creating connections between them, is the mind mapping method~\cite{Bates_2019}. Figure~\ref{fig1} illustrates a mind map of strategies that students can follow to increase attentiveness in the classroom. Some key strategies that students need to embrace to improve attentiveness are taking a digital detox, minimizing digital distractions, taking effective notes, actively participating in class, learning autonomy, properly managing tasks, and following a healthy lifestyle. The central concept of increasing attentiveness sits at the center and branches outwards to the six listed strategies, as shown in Fig.~\ref{fig1}. The secondary, tertiary, and quaternary branches can be created, extending the key point to add sub-subtopics and details. Each strategy in Fig.~\ref{fig1} further branches outwards, connecting different approaches students may want to follow. For example, the notetaking strategy connects tactics of having handwritten notes instead of digital or online application written notes, adding visual concepts, using memory aids, preparing structurally consistent notes, and having mnemonics, symbols, and abbreviations instead of writing complete sentences or paragraphs.   
 
A digital detox helps reduce stress by alleviating cognitive overload, improve memory, enhance focus, and allow the brain to rest and process information more effectively, thereby improving mental clarity~\cite{KoRa_2024}. Moreover,  the digital detox helps in students' psychological development and improvement of students' concentration~\cite{HasHam_2025}. 
Although the use of digital technology is unavoidable in the current digital transformation era, a brief discussion on students’ necessity of taking a digital detox during classes to increase attentiveness is in order. Unless the learning strategy, instructional approach, and subject and content-like contextual factors demand, a wise choice for the students to improve attentiveness in class is not to use network and internet-connected electronic devices such as cell phones, smartwatches, laptops, tablets, etc., during lectures, which not only distract students but also negatively impact in students’ participation in classroom activities. Students are strongly recommended to turn off notifications during class and follow daily communication management strategies such as setting disconnection time, no screen time, communication preferred time, time for social media interaction, and decreasing internet addiction, etc. 
Mind mapping diagrams are free-flowing and helpful for brainstorming, finding linkage between ideas/concepts/topics, and generating ideas, keeping the main topic at the center. One may choose to prepare or recommend students to create concept maps as they are more structured, focused on connecting different concepts, and follow a hierarchy.

Many other effective notetaking methods are also in practice such as the charting method, outline method, sentence method, flow method, boxing method, question-evidence-conclusion (QEC) method, split page method, writing on slides, bullet point method, list method, and rapid logging method~[40-46].\nocite{gold_1918_outline, Hawley_1922_sentence, Numazawa_2016, Kiewra_1991, Alda_2024, Newport_2007, Carroll_2018_bullet}  The choice of notetaking method depends on many factors such as learning style, subject matter and topic, lecture structure and content delivery method, context, learner's personal goal and preferences, available time to prepare notes, and need for mathematical, logical, and visual sophistication.

\section{\label{sec:level3}   Methods}

\subsection{Class and Lecture Design}
The courses were designed based on the backward design of effective teaching, prioritizing the intended learning outcomes and course objectives over the chapters to be covered in the course, as well as the choice of instructional methods and assessment types.  The backward design instructional method is based on aligning instruction, course content, pedagogy, and assessment with the learning objective, in contrast to the traditional teaching method of starting the learning process based on teacher-designed content, delivering lectures, assigning problems, and assessing students based on what was taught~\cite{WiMc2005}. In the backward design method, one first determines the desired learning objectives and goals that students are expected to learn by the end of the day, chapter, and the course. Second, assessments are determined and planned to check whether students have learned what they were expected to learn. Third, learning activities and tasks are prepared that will be done in the classroom and beyond to prepare students for assessments. Algebra-based physics is primarily designed for students majoring in life sciences rather than students majoring in physical science. Students pursuing careers in life sciences are often less motivated and less prepared for physics courses, which require active, goal-oriented learning rather than passive, teacher-centered methods. To address this, we used a backward design approach in teaching these courses. 

 The classes met twice a week for 75 minutes each day. Pedagogy includes lecturing, simulations, demonstrations, problem-solving, hands-on activities, think-pair-share, and class work as multiple choice questions. The classwork covered topics discussed at the beginning and end of the lecture, which helped ensure that students arrived on time and stayed for the whole class and boosted class attendance~\cite{Weinstein_Book_2018}.  Students were strongly recommended to take a digital detox in the classroom, which minimizes digital distractions and helps minimize multi-tasking, which is considered one of the major learning pitfalls~\cite{Madore_Nature_2020}.
Students using electronic devices to take notes are prone to multitasking, but multitasking during learning impacts comprehension and knowledge retention as attention gets divided, focus decreases, the brain needs to pay the task-switching costs on refocusing, the brain receives cognitive load usually more than it can handle with no stress, and working memory is disrupted~\cite{Jamet_2020_EduSci}. 
The assignments include online quizzes, primarily focused on concept development, and instructor-prepared homework assignments focused on problem-solving. The instructor used a liquid syllabus~\cite{Murphy_2023}, which is a dynamic, web-based, interactive, and student-engaging syllabus that includes visual elements such as figures and videos, links to resources, and QR codes. The liquid syllabus has fixability in terms of dates, learning activities, and so on, to ensure the course objective is met by the end of the semester.  The liquid syllabus, featuring more multimedia components, was adopted, inspired by the fact that students rarely read the syllabus once at the beginning of the semester and need to be made aware of the instructors’ expectations, requirements, and core objectives for the content and the course. A study shows that a liquid syllabus provides students with what they need to succeed in the course, increases equity, fosters a sense of belonging and respect through student-friendly language and visuals, and offers a humanizing introduction for the course~\cite{Pacansky_2021}. Students must complete one prerequisite course before enrolling in Physics I. At Fayetteville State University (FSU), students must have a minimum passing grade in one of 4 math courses: college trigonometry, algebra and trigonometry, pre-calculus mathematics I, or pre-calculus mathematics II.

\begin{table}[htb!] 
\renewcommand{\arraystretch}{1.25}
\caption{FSU's University catalog to assign letter grade for earned grade points on a scale of 0--100 and a GPA out of 4.0.\label{tab1}}
\centering
 \begin{tabular}{|c| c| c| c| c |c|} 
 \hline
Points& 90 -- 100	& 80 -- 89.9 &  70 -- 79.9 &  60 -- 69.9 & < 60\\
\hline
Grade& A & B& C & D & F\\
\hline
GPA & 4.0 &  3.0  & 2.0 & 1.0 & 0.0\\
\hline
\end{tabular}
\end{table}
%


\begin{figure}[htb!]
\centering
\includegraphics[width= 7.5 cm]{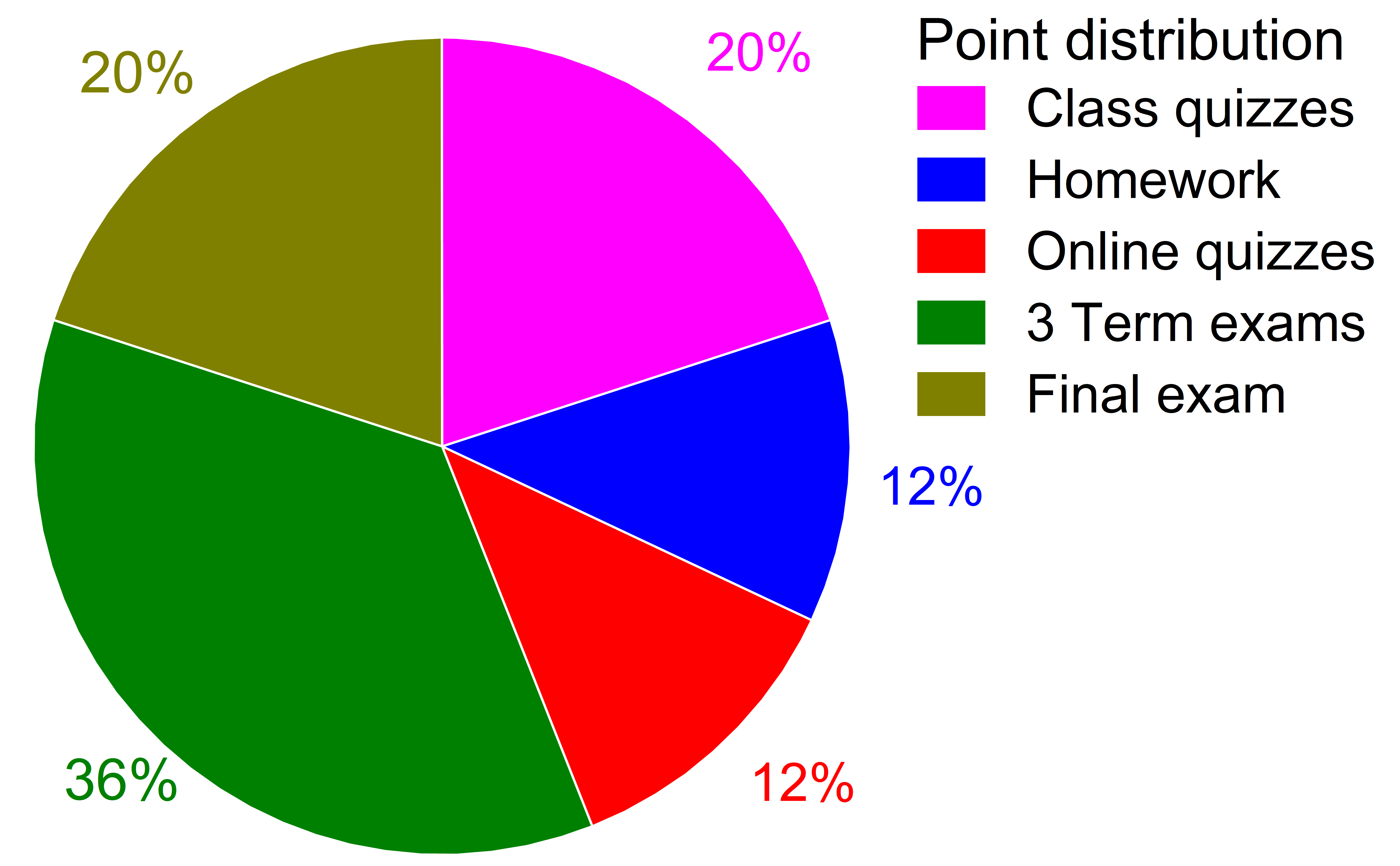}
\caption{ A Pie chart showing point distribution in introductory physics course. Exams weighs 56\% towards the final grade. Each mid term exam has 12\% weight. \label{pieC}}
\end{figure}
The grading practices outlined in Table~\ref {tab1} were communicated at the beginning of the semester and a couple of times thereafter. The grade points are made student-trackable to ensure students understand the grading practices, can identify their strengths and weaknesses, and stay focused on their goals. Assignments were structured to \textit{(i)} reduce the negative impact of late work, keeping students responsible for submitting assignments on time, \textit{(ii)} give chances to implement feedback for improvement, and \textit{(iii)} provide opportunities for students to reflect on and improve their performance, which motivated student in learning by doing and revising. We focused on reducing the basis of grading by emphasizing academic achievement and excellence, excluding non-academic components such as attendance, seminar participation, group discussion, extra work for points, etc.
Students' final grade was determined by aggregating their scores across assignments, with each assignment weighted according to its assigned weight. Online tests were provided with three possible attempts for each question, in which a link to the relevant section in the eTextbook and available resources was provided.   A hint with insight into how to answer the questions was made visible after the first attempt, and the detailed solutions were provided a minute after the deadline.  Numerical values were randomized for each student, and the order of answer choices was shuffled in each attempt. The desired number of significant figures was made visible for each question. Each student got feedback on each attempt. The best score out of the maximum three attempts was recorded.  Students are encouraged to increase attentiveness, as we presented in Fig.~\ref{fig1}, and be active physically and mentally in the learning environment rather than just being passive listeners.

Recognizing that elaborated rehearsal, implementation of feedback, repetition of tasks, and gaining overall skills are crucial, we assigned a lower weight to problem-solving homework assignments and online quizzes than terminal exams and the final comprehensive exam. For the same reason, the final exam had a slightly larger weight than each terminal exam.   A schematic of point-distribution is presented in a pie-chart in Fig.~\ref{pieC}. The pie chart is presented only for reference. The students' grade presented in this study to analyze students' performance is taken only from the third mid-term exams, which is the last mid-term before the final exam. The midterm exams were the instructor-prepared questionnaires. The problems were mostly problem-solving, analytical analysis, and critical thinking rather than knowledge transfer.  The third mid exam (analyzed here)  covered two chapters, namely (1) work and energy and (2) Impulse and momentum.  Thus, the grade presented in this study is not the final grade that students received; rather, it is a 12\% contribution towards their final grade.

\subsection{Study Design}
FSU follows the FN grade policy, which means that if students do not attend the first week of class, they will automatically drop out after the instructor notifies the registration office.  There was no mid-semester withdrawal. Two students in the sections implementing no notetaking and one student from sections where notetaking was implemented did not take the midterm exam. Students dropped due to FN grades, and those who did not take the exam are not included in the analysis.
Keeping the other factors the same,  teaching pedagogy differing with the preparation and use of notetaking in two sets of sections was implemented. Instructor taught effective notetaking methods, asked students to prepare handwritten notes, and let them use the notes during classwork, timed quizzes, and exams. Study shows that hand-written notes increase comprehension, create a personal connection to the learning materials, improve memory and recall of the facts, and activate brain memory areas more robustly than preparing notes using digital devices, showing the pen mightier than the keyboard, although taking notes on electronic devices is faster, and the note taker can write to a larger extent of word-for-word what the teacher is saying in a class~\cite{Muller_PsySci_2014}.  Students using electronic devices such as iPads, tablets, and laptops to take notes are usually prone to copying, pasting, and collecting information without a deep thought about what they are writing.  In contrast, hand-written note-takers usually mentally process the information further before they write it down; while they write, they get their fingers’ extra muscle memory. 
Therefore, this research is designed as a comparative study of the effectiveness of handwritten notes in enhancing attention and effectiveness in physics learning. In order to maintain consistency, the quizzes and exams administered for both groups were problem-solving based, which demand critical thinking, creativity, and decision-making. Keeping all other conditions and class structure unaltered, the methodology was split into two main categories.
\begin{enumerate}
    \item Handwritten notes are prohibited in exams; instead, a list of equations is provided:
    \begin{itemize}
        \item[$-$] This policy was implemented in the first introductory physics lecture course, where the notetaking strategy was not discussed, nor were students allowed to use their notes in the exams. However, a list of all required mathematical expressions covered in the chapters, aka the formula sheet, was provided during timed assignments and exams.  Scores from a terminal (third) exam are analyzed.
  \end{itemize}
    \item Handwritten notetaking is encouraged and permitted for use during exams:
         \begin{itemize}
        \item[$-$] The notetaking policy was implemented in the first introductory physics course. At the beginning of the semester, a few samples of the notetaking skeleton were provided, and instructors consistently observed students' notes during class to assess their effectiveness and guided them as needed. Only the handwritten notes prepared by students were permitted during exams. The notes students used in the timed quizzes and exams were monitored.  The scores students received in a terminal (third) exam are analyzed.
    \end{itemize}
\end{enumerate}

In the sections where the notetaking strategy was implemented, a few samples of skeleton notes of each kind discussed above, along with sample notes, were also provided in the very first week of the semester. In addition, students received timely feedback on their notes, and those who struggled with notetaking and other academic things were encouraged to use office hours for additional support. The instructor had at least eight regularly scheduled office hours per week to help students with various aspects, including course content, academic and career plans, and other academic support. The instructor asked students to make mind maps connecting different concepts in all the chapters covered as notetaking homework, which the instructor found very helpful for students' overall understanding of the course.

\section{\label{sec:level5}   Results and discussion }

The grade that each student receives is the grade that students earn based on their academic performance, not measured against the performance of other students. The letter grades are assigned according to the University's catalog as  tabulated in Table~\ref{tab1}. At FSU, using the traditional five-grade system, a letter grade of "A" (GPA of 4.0 out of 4.0) is assigned if a student receives 90\% of possible credits or more than that. Letter grades "B" (GPA of 3.0 out of 4.0), "C" (GPA of 2.0 out of 4.0), and "D" (GPA of 1.0 out of 4.0) are assigned if a student's grade point lies in the range of 80 to 89.9\%, 70 to 79.9\%, and  70 to 79.9\%, respectively. A student fails the course if the student receives less than 60\% of the total possible points.  

\begin{figure}[tbh!]
\centering
\includegraphics[width=7.5 cm]{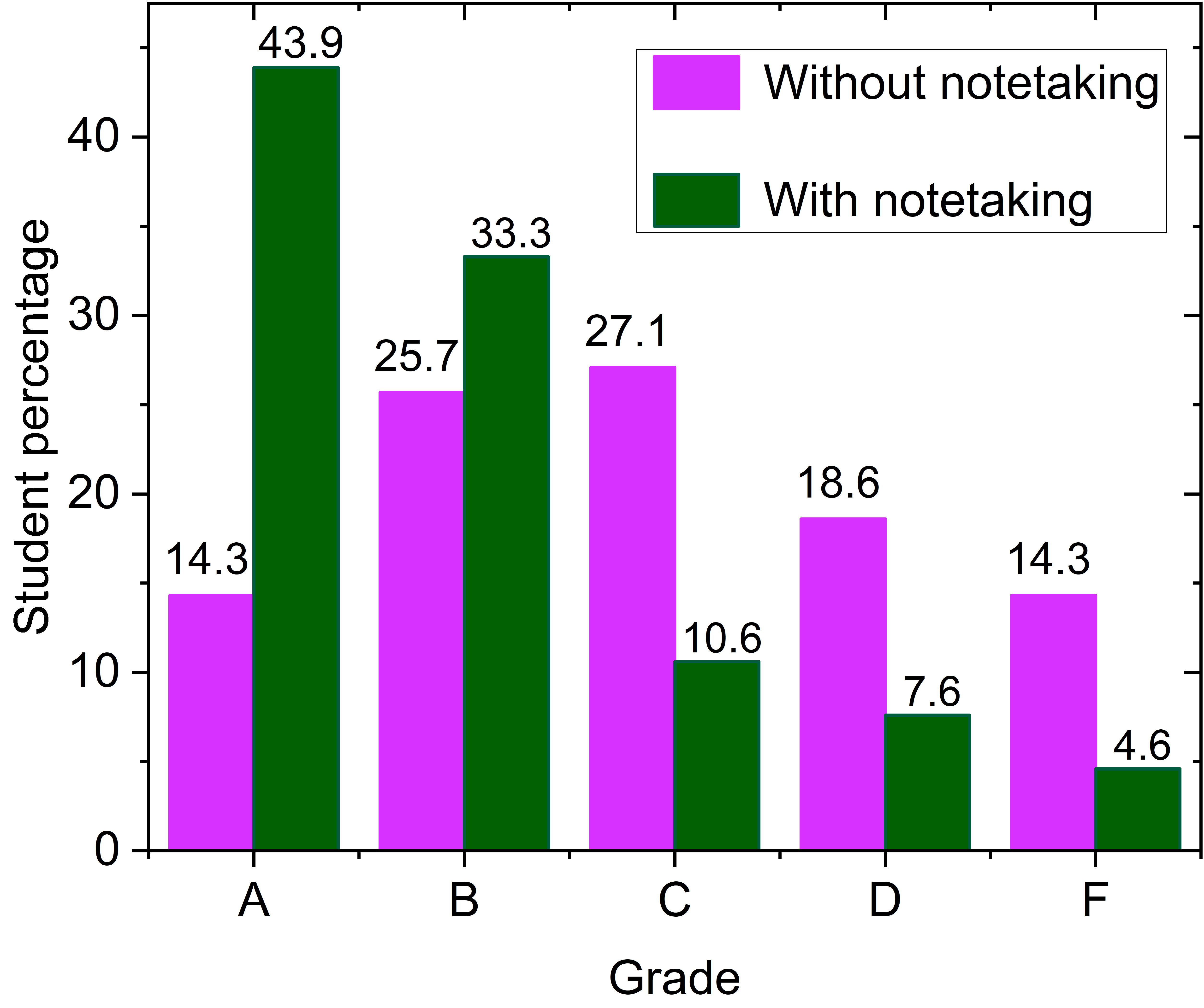} 
\caption{Bar chart showing letter grade percentage in sections with no notetaking implemented and with notetaking implemented.  The grade presented in this bar chart is not the final grade student received, instead it is the grade students received in a mid-term exam.  There were a total of 69 students involved in the teaching approach with no notetaking and total of 66 students in the notetaking implemented group.\label{fig23}}
\end{figure}

\begin{table}[htb!] 
\renewcommand{\arraystretch}{1.25}
\caption{Statistical measures of students' performance in two different approaches to teaching introductory physics course in a terminal exam. The highest, lowest, and average scores were presented on the scale of 0--100.\label{tab:table3}}
\centering
 \begin{tabular}{|c| c| c|} 
 \hline
  Statistical measures    & Without notetaking &   With notetaking \\
\hline
Sample size & 69 		& 66 \\
 \hline
  Highest score                & 100       & 100 \\
\hline
  Lowest   score              & 31        & 55   \\
\hline
  Average   score             & 75.3      & 85.4     \\
\hline
  Standard deviation     & 14.4       & 11.0   \\
\hline
\end{tabular}
\end{table}

Figure~\ref{fig23}  shows the percentage of letter grades that students received in sections with no notetaking and sections with notetaking were implemented. The sample size for these two groups in the study is almost the same, one being 69 and the other being 66.
The data shows that when students were allowed to take notes, the percentage of students receiving the highest letter grade (i.e., "A")
increased significantly, the second-highest letter grade (i.e., "B") increased slightly, while the lower letter grades decreased significantly.
The percentage of students getting the letter grade "A" is more than triple with the implementation of notetaking (please see and compare the pair of bars for letter grade "A"). More than 40\% of students mastering the content and getting "A" grades indicate that the notetaking strategy works well in learning physics.  Students receiving letter grades "C" and "D" and failing grade percentages dropped significantly when notetaking was implemented.  Without notetaking, the letter grade peaked at "C" while the letter grade peaked at "A" and tailed to a lower grade when notetaking is implemented. The percentage of students who failed the course decreased from 14.3\% to only 4.6\%.

The statistical measures of student performance while the two different notetaking policies were implemented in teaching are presented in Table~\ref{tab:table3}. Two out of 69 and one out of 66 students received perfect scores in the no-notetaking and with-notetaking groups, respectively. The average score in the class improved significantly ($\sim 10\%$) in the sections that used handwritten notes, although there was a lowest score outlier of 31. The standard deviation, which measures the sparseness of scores about the average score, decreased in the sections where the notetaking policy was implemented, indicating overall performance improvement in those sections.
The findings of this study, concluding that effective notetaking helps in academic excellence, are consistent with those of other studies examining the role of effective notetaking in learning.  This study also affirms the concept of cognetive load theoy that instructional methods can be optimized to improve understanding and  retain knowledge longer~\cite{de_InstrSCI_2010}.

\section{Statistical Tests and Limitation of the Study}

At a significance level of $\alpha=0.05$ (or 95\% confidence interval), the statistical power is about 70\%, meaning there is a 70\% chance of correctly detecting a true effect. In statistical hypothesis testing, this is known as a Type II error, also called a false negative~\cite{Dekking_2005}.  To reliably reject a false null hypothesis, a statistical power of 80\% or higher is commonly recommended at a significance level of  $\alpha=0.05$~\cite{Cohen_1988}.
\begin{figure}[h!]
\centering
\includegraphics[width= 7.5 cm]{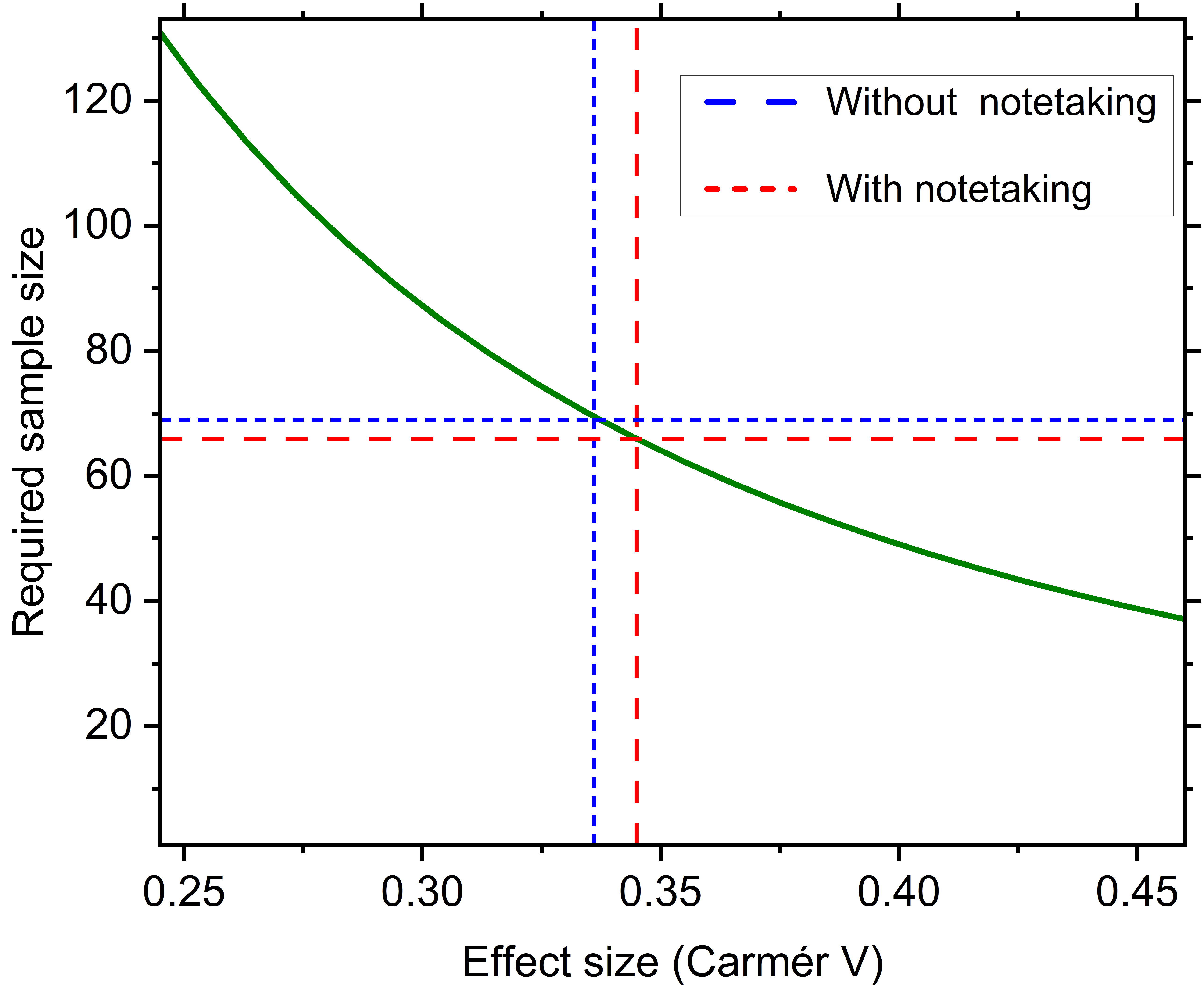}
\caption{ Plot showing the required sample size for 80\% statistical power for as a function of effect size in terms of Cramér’s $V$-values. \label{fig4}}
\end{figure}
To gain a better understanding of the results, one can evaluate the effect size using measures like Cramér’s V, which can be calculated using the following formula:
\begin{equation}
V = \sqrt{\frac{\chi^2}{n \cdot (k - 1)}}\,,
\end{equation}
where $\chi^2$, $n$ and $k$ are chi-square statistic,  total number of observations and number of categories respectively.  As shown in Fig.~\ref{fig4}, the Cram\'er’s $V$ values for the sample size of 69, i.e, without note taking, is 0.335, while that for the sample size of 66, i.e., with note taking, is 0.345.  The $\chi^2$ values of students' scores in the group without notetaking is 8.90 with 4 degrees of freedom.  The same for notetaking implemented group is 35.52 with 4 degrees of freedom.  The Cram\'er’s $V$ value for the dataset for the notetaking not implemented group is 0.180, showing a weak association and suggesting that students' score distribution differs only slightly from the uniform distribution. On the other hand, Cram\'er’s $V$ value for the notetaking implemented group is 0.367, indicating a moderate association strength and indicating scores concentrated in the higher bins 80-90 and 90-100, i.e., higher letter grades.
The study was conducted at only one institution offering a limited number of introductory physics course sections. Expanding the study to include multiple institutions and a larger student population would enhance the reliability of the results. Reliability could be  improved by  conducting the study with half of the students in a class using notetaking strategies, while the other half do not, in which students could self-select their group or be randomly assigned by the instructor. Alternatively, one can investigate the role of notetaking by implementing the above-mentioned strategies to assess the same group of students across different midterm exams. Furthermore, FSU is a historically black university in which the majority of students enrolled are from historically underserved groups. Carrying out the same or similar study in a diverse population would increase reliability of the research findings.

We acknowledge the fact that “grades” are considered the outcome. In many contexts, grades are short-term measures, while conceptual retention is a long-term measure. Conceptual retention over the years is out of the scope of this study. One may want to study the effect of notetaking measures in various other settings, such as larger universities, community colleges, and non-STEM courses, and compare the outcomes with those presented in this study. 

 \section{ Conclusion}

This study investigates the impact of effective notetaking on learning outcomes in undergraduate-level  introductory physics courses at a minority-serving university. 
We highlighted  the importance of a distraction-free, student-centered active learning environment, which enhances the retention of learning materials in long-term memory. This approach enables students to retrieve and apply the knowledge, concepts, and skills that they have acquired in various contexts both now and in the future.
A few commonly used notetaking methods were discussed. As an example of the mind mapping method of notetaking, we have presented a mind map to discuss techniques that students may follow to increase their attentiveness in the classroom, making physics learning fruitful.  
We discussed various approaches we have implemented in teaching and assessing algebra-based undergraduate-level introductory physics courses, as well as the impact of our efforts to make physics learning fun, active, student-centered, interactive, and objective-oriented.  

We found that the majority of students were motivated to prepare better notes and actively engage in the classroom, as they could use their handwritten notes in timed assignments and exams. 
The letter grades that students received for a mid-term exam were recorded as a quantitative measurement of learning effectiveness.
The letter grade significantly improved when students were allowed to prepare and use the notes they prepared during exams and timed assignments. Not only did the percentage of students getting higher grades increase with notetaking implementation, but the percentage of students failing the course also decreased sharply. 

The findings recommend that educators consider effective notetaking as an integral part of learning activities.
In today's digital age, minimizing distraction is crucial.
Asking students to make their mind map connecting different concepts in all the modules (chapters) covered in a course as notetaking homework, which we encouraged students to prepare, is an excellent homework any instructor can assign. 

\section*{Institutional Review} 
This study was reviewed and approved by the Human Rights in Research Committee (HRRC) at Fayetteville State University with the assigned  IRB \#2026-20 under "Exempt Status".

\begin{acknowledgments}
The author acknowledges the communications with Drs. Tikaram Neupane and Uma Poudyal. 
\end{acknowledgments}


\begin{thebibliography}{999}
\bibitem[Atkinson(1971)]{Atkinson_SciAm_1971}
Atkinson, R.C.; Shiffrin, R.M. The Control of Short-Term Memory. {\em Sci. Am.} {\bf 1971}, {\em 225}, 82–90.
%
\bibitem[ Balduccini(2011)] {Balduccini_book_2011}
Balduccini, M.; Girotto, S. ASP as a Cognitive Modeling Tool: Short-Term Memory and Long-Term Memory. In {\em  Logic Programming, Knowledge Representation, and Nonmonotonic Reasoning: Essays Dedicated to Michael Gelfond on the Occasion of His 65th Birthday};   Balduccini, M.; Son, T.C. Eds.; Springer, 2011; pp. 377–397.
%
\bibitem[Felder(2024)]{Felder__book_2016}
Felder, R. M.;  Brent, R. \textit{Teaching and Learning STEM: A Practical Guide}, Jossey-Bass: San Francisco, CA, USA, 2016; pp. 111--133.
%
\bibitem[Chew(2021)]{Chew_CanPsychol_2021} 
Chew, S. L. An advance organizer for student learning: Chokepoints and pitfalls in studying. {\em Can. Psychol./Psychol. Can.} {\bf 2021}, {\em 62}, 420.%
%
\bibitem[SWeller(2016)]{Sweller_2016JJarMac}
Sweller, J.  Working memory, long-term memory, and instructional design. {\em J. Appl. Res. Mem. Cogn.} {\bf 2016}, {\em 5}, 360. https://doi.org/10.1016/j.jarmac.2015.12.002.
%
%
\bibitem[Jin(2024)]{Jin_SLA_2024}
Jin,  Z. and Webb, S. The effectiveness of note taking through exposure to L2 input: A meta-analysis. {\em Studies in Second Language Acquisition}, {\bf 2024}, {\em 46}(2), 404--426. doi:10.1017/S0272263123000529 
%
\bibitem[Jin(2021)]{Jin_Webb_2021}
Jin, Z. and Webb, S. Does writing words in notes contribute to vocabulary learning? {\em Language Teaching Research} {\bf  2021}, {\em 0}, 136216882110621. https://doi.org/10.1177/13621688211062184
%
\bibitem[Lee(a998)]{Lee_Thesis_1998}
Wilberschied, L. F. The relationships among a variation of immediate recall tasks and measures of L1 writing, L2 achievement, and cognitive strategy use in students of high school Spanish, Doctoral Dissertation, The Ohio State University, Columbus, Ohio, USA, ({\bf 1998}).
%
\bibitem[Campana(2009)]{Campana_Thesis_2009}
Campana, J. N. The Effectiveness of Using Guided Notes In a High School Physics Classroom, MS Thesis, St. John Fisher University, Rochester, New York, USA ({\bf 2009}). 
%
\bibitem[Anastasiou(2024)]{AnastasiouEtal_EPR_2024} Anastasiou, D.; Wirngo, C.N.; Bagos, P. The Effectiveness of Concept Maps on Students’ Achievement in Science: A Meta-Analysis. {\em Educ. Psychol. Rev.} {\bf 2024}, {\em 36}, 39. https://doi.org/10.1007/s10648-024-09877-y.
%
%
\bibitem[Shi(2022)]{Shi_Etal_AJET_2022} Shi, Y.; Yang, H.; Yang, Z.; Liu, W.; Wu, D.; Yang, H. H. Examining the effects of notetaking styles on college students’ learning achievement and cognitive load. {\em Australas. J. Educ. Technol.} {\bf 2022}, {\em 38(5)}, 1–11. https://doi.org/10.14742/ajet.6688
%
\bibitem[Fanguy(2025)]{Fanguy_AJET_2025} Fanguy, M.  Beyond traditional notetaking: Comparing the effects of collaboratively-generated versus instructor-provided notes on recall and writing performance.   {\em Australas. J. Educ. Technol.} {\bf 2025}, {\em 41(1)}, 1–17. https://doi.org/10.14742/ajet.9446.
%
\bibitem[Weinstein(2018)]{Weinstein_Book_2018} 
Weinstein, Y.; Sumeracki,  M.; Caviglioli, O. \textit{Understanding how we learn: A visual guide}, Routledge: New York, NY, USA, 2019.
%
\bibitem[Voyer(2022)]{Voyer_2020_CEP}
Voyer, D.; Ronis, S. T.; Byers, N.  The effect of notetaking method on academic performance: A systematic review and meta-analysis. {\em  Contemp. Educ. Psychol.} {\bf 2022}, {\em 68}, 102025. https://doi.org/10.1016/j.cedpsych.2021.102025
\bibitem[Bohay(2011)]{Bohay_2011_AJP}
Bohay, M.; Blakely, D. P.; Tamplin, A. K.; Radvansky, G. A. Note Taking, Review, Memory, and Comprehension. {\em Am. J. Psychol.} {\bf 2011}, {\em 124}(1), 63–73. https://doi.org/10.5406/amerjpsyc.124.1.0063
%
\bibitem[Jaleel(2024)]{JaleelThomas2024}
Jaleel, S.; Thomas, A. M. \textit{Learning Styles: Theories and Implications for Teaching Learning}, Horizon Research Publishing: San Jose, CA, USA, 2024.
%
\bibitem[Bain(2024)]{Bain_Book_2023}
Bain, K.\textit{What the Best College Students Do}, Harvard University Press: London, England, 2012.
%
\bibitem[Salame(2024)]{Salame_IJI_2024}
Salame, I. I.; Tuba, M.; Nujhat, M. notetaking and Its Impact on Learning, Academic Performance, and Memory. {\em International Journal of Instruction} {\bf 2024}, {\bf 17}(3), 599–616.
%
\bibitem[Georgakis(2017)]{Georgakis_2017}
Georgakis, A.\textit{How To Take Good Notes: The Science Behind notetaking},  CreateSpace Independent Publishing Platform: Scotts Valley, California, USA, 2017.
%
\bibitem[Jones(2015)]{Jones_2015}
Jones,B. \textit{Note Taking: 10 Simple Steps to Effective Note Taking}.United States, CreateSpace Independent Publishing Platform: Scotts Valley, California, USA, 2015.
%
\bibitem[Artz(2020)]{Artz_JEC_2020}
Artz, B.; Johnson, M.; Robson, D.; Taengnoi, S. Taking notes in the digital age: Evidence from classroom random control trials. {\em Journal of Economic Education}, \textbf{2020}, {\em 51}(2), 103–115. https://doi.org/10.1080/00220485.2020.1731386
%
\bibitem[Boye(2012)]{Boye_2012}
Boye, A. notetaking in the 21st century: Tips for instructors and students {\em [White paper]}. Texas Tech University. {\bf 2012}.
%
\bibitem[Cohen(2013)]{Cohen_2013}
Cohen, D.; Kim, E.; Tan, J.; Winkelmes, M.-A. A note-restructuring intervention increases students’ exam scores. {\em College Teaching}, {\bf 2013}, {\em 61}(3), 95--99.
%
\bibitem[Dynarski(2017)]{Dynarski_2017}
Dynarski, S.  For note taking, low-tech is often best. {\em Harvard Graduate School of Education}. (August 21, 2017).
%
\bibitem[Gonzalez(2018)]{Gonzalez_2018}
Gonzalez, J.  notetaking: A research roundup. {\em Cult of Pedagogy}. 
Available online: https://www.cultofpedagogy.com/notetaking/ (accessed on 12 September 2024).
%
\bibitem[Kauffman(2011)]{Kauffman_2011}
Kauffman, D. F.; Zhao, R.; Yang, Y.-S.  Effects of online note taking formats and self-monitoring prompts on learning from online text: Using technology to enhance self-regulated learning. {\em Contemporary Educational Psychology} {\bf 2011}, {\em 36}(4), 313–322.
%
\bibitem[Mueller(2018)]{Mueller_2018}
Mueller, P. A.; Oppenheimer, D. M. Corrigendum: The pen is mightier than the keyboard: Advantages of longhand over laptop note taking. {\em Psychological Science} {\bf 2018}, {\em 29}(9), 1565–1568. https://doi.org/10.1177/0956797618781773
%
\bibitem[Rahmani(2011)]{Rahmani_2011}
Rahmani, M.; Sadeghi, K.  Effects of notetaking training on reading comprehension and recall. {\em The Reading Matrix} {\bf 2011}, {\em 11}(2), 116--128.
%
\bibitem[Reynolds(2016)]{Reynolds_2016}
Reynolds, S. M.; Tackie, R. N.  A novel approach to skeleton-note instruction in large engineering courses: Unified and concise handouts that are fun and colorful [Paper presentation]. American Society for Engineering Education Annual Conference \& Exposition, New Orleans, LA, USA (26-29 June 2016). https://www.asee.org/public/conferences/64/papers/15115/view
%
\bibitem[Robin(1977)]{Robin_1977}
Robin, A.; Foxx, R. M.; Martello, J.; Archable, C. Teaching notetaking skills to underachieving college students. {\em The Journal of Educational Research} {\bf 1977}, {\em 71}(2), 81--85. https://doi.org/10.1080/00220671.1977.10885042
%
\bibitem[Stutts(2013)]{Stutts_2013}
Stutts, K. J.; Beverly, M. M.; and Kelley, S. F. Evaluation of note taking method on academic performance in undergraduate animal science courses.{\em NACTA Journal} {\bf 2013}, {\em 57}(3), 38--39.
%
\bibitem[Wu(2018)]{Wu_2018}
Wu, J. Y.; Xie, C.  Using time pressure and notetaking to prevent digital distraction behavior and enhance online search performance: Perspectives from the load theory of attention and cognitive control. {\em Computers in Human Behavior}, {\bf 2018}, {\em 88}, 244--254. https://doi.org/10.1016/j.chb.2018.07.008
%
\bibitem[Pauk(2010)]{Pauk_2010}
Pauk, W.; Owens, R. J. Q. The Cornell System: Take Effective Notes. In {\em How to Study in College  (10 ed.)}; Cengage Learning: Boston, MA, USA, 2010; pp. 235--277. 
%
\bibitem[Bui(2015)]{Bui_JARMC_2015}
Bui, D. C.;  McDaniel, M. A. Enhancing learning during lecture notetaking using outlines and illustrative diagrams, {\em J. Appl. Res. Mem. Cogn.},
{\bf 2015}, {\em 4}(2), 129--135. https://doi.org/10.1016/j.jarmac.2015.03.002.
%
\bibitem[Gupta(2024)]{Gupta_2024}
Gupta, R.; Sangeetha, A., Keni, G. K. \textit{The Roadmap for Academic Success: Essential Student Strategies for Time Management, Focus, Beating Procrastination, Learning, and Memory}, Notion Press, Chennai, India, {\bf 2024}.
%
\bibitem[Kadavy(2021)]{Kadavy_2021}
 Kadavy, D· \textit{Digital Zettelkasten: Principles, Methods, \& Examples}. Kadavy Inc.:Walnut, CA, USA, {\bf 2021}.
 %
\bibitem[Bates(2019)]{Bates_2019}
Bates, T. \textit{How to mind map: 7 easy steps to master mind mapping techniques, notetaking, creative thinking \& brainstorming skills}. Lulu. com, {\bf 2019}.
%
\bibitem[Kour(2024)]{KoRa_2024} Kour, J.; Rathee, N. Importance of digital detox for cognitive rejuvenation. 
\textit{Human Cognition: In the Digital Era}; Singh, D.; Uniyal, S., Eds,; Clever Fox Publishing, Chennai, India, 2024; pp. 13--21.
%
\bibitem[Hasanah(2025)]{HasHam_2025} Hasanah, R.; Hamdi, T. T. Digital Detox in Schools: Balancing Technology Use with Students' Psychological Development. {\em J. Educ. Manag. Res.} {\bf  2025}, {em 1}, 1--12. https://ejournal.bospintar.com/index.php/Je/article/view/7
%
\bibitem[gold(1918)]{gold_1918_outline}
Gold, M. S.  Practical Suggestions for the History Teacher: The Use of the Outline Method in the Teaching of History {\em Historical Outlook} {\bf 1918}, {\em 9}(7), 381--383.
%
\bibitem[Hawley(1922)]{Hawley_1922_sentence}
Hawley, W. E.  The “List” versus The “Sentence” Method of Teaching Spelling. {\em The Journal of Educational Research} {\bf1922}, {\em 5}(4), 306--310. https://doi.org/10.1080/00220671.1922.10879257
%
\bibitem[Numazawa(2016)]{Numazawa_2016}
M. Numazawa; M. Noto, The effect of education and learning using notetaking application.  IEEE International Conference on Systems, Man, and Cybernetics (SMC), Budapest, Hungary, 9-11 October 2016. doi: 10.1109/SMC.2016.7844345.
%
\bibitem[Kiewra(1991)]{Kiewra_1991}
Kiewra, K. A.; DuBois, N. F.; Christian, D.; McShane, A.; Meyerhoffer, M.; Roskelley, D.  notetaking functions and techniques. {\em Journal of Educational Psychology} {\bf 1991}, {\em 83}(2), 240--245. https://doi.org/10.1037/0022-0663.83.2.240
%
\bibitem[Alda(2024)]{Alda_2024}
Alda, R. P.  Exploring Student’s notetaking Strategies In Reading English Text. Bachelor of Education Thesis, Faculty of Education and Teacher Training,  Ar-Raniry State Islamic University, Banda Aceh, Indonesia, {\bf 2024}.
%
\bibitem[Newport(2007)]{Newport_2007}
Newport, C. \textit{How to Become a Straight-A Student: The Unconventional Strategies Real College Students Use to Score High While Studying Less}. Crown Publishing Group: New York City, NY, USA, {\bf 2007}.
%
\bibitem[Carroll(2018)]{Carroll_2018_bullet}
Carroll, R. \textit{The Bullet Journal Method Collector's Set}, Penguin Publishing Group: City of Westminster, London, {\bf 2018}.
%
\bibitem[Wiggins(2005)]{WiMc2005}
Wiggins, G. P.; McTighe, J. \textit{Understanding by design}, expanded 2nd ed.; Association for Supervision and Curriculum Development, 
Alexandria, VA, USA, 2005. 
%
\bibitem[Madore(2020)]{Madore_Nature_2020}
Madore, K.P.; Khazenzon, A.M.; Backes, C.W.; Jiang, J.; Uncapher, M. R.;  Norcia,  A. M. and Wagner, A. D. Memory failure predicted by attention lapsing and media multitasking. {\em Nature} {\bf 2020}, {\em 587}, 87--91. https://doi.org/10.1038/s41586-020-2870-z
%
\bibitem[Jamet etal(2020)]{Jamet_2020_EduSci} Jamet, E.; Gontihier, C.; Cojean, S.; Colliot, T.; Erhel, S. Does multitasking in the classroom affect learning outcomes? A naturalistic study. {\em Comput. Hum. Behav.} {\bf 2020}, {\em 106}, 106264. https://doi.org/10.1016/j.chb.2020.106264
%
\bibitem{Murphy_2023}
Murphy-Bowne, M. J. \textit{A Liquid Syllabus: A Visual Starting Point}. In J. Lee, W. Huang, X. Chen, F. Rodrigues, L. Okan, S. Beene, C. Huilcapi-Collantes (Eds.),{\em Connecting \& Sharing: The Book of Selected Readings 2023}; The International Visual Literacy Association; 2023; pp.128-139.
https://doi.org/10.52917/ivlatbsr.2023.019 
%
\bibitem[Pacansky(2021)]{Pacansky_2021}
Pacansky-Brock, M.  The liquid syllabus: An anti-racist teaching element. {\em Colleague 2 Colleague Magazine} {\bf 2021}, {\em 1}(15), 2.
%
\bibitem[Muller(2014)]{Muller_PsySci_2014}
Mueller, P. A.; and  Oppenheimer, D. M. The Pen Is Mightier Than the Keyboard: Advantages of Longhand Over Laptop Note Taking. {\em Psychological Science} {\bf 2014}, {\em 25}(6), 1159--1168.
https://doi.org/10.1177/0956797614524581
%
\bibitem[Dekking(2005)]{Dekking_2005}
Dekking, F. M.;  Kraaikamp, C.; Lopuha\"{a}, H. P.;  Meester, L. E. \textit{A modern introduction to probability and statistics : Understanding why and how}. Springer-Verlag London Ltd.,  London, {\bf 2005}. https://doi.org/10.1007/1-84628-168-7
%
\bibitem[Cohen(1988)]{Cohen_1988}
Cohen, J.  \textit{Statistical Power Analysis for the Behavioral Sciences}. 2nd ed. Routledge, New York, USA {\bf 1988}.  \\
https://doi.org/10.4324/9780203771587
%
\bibitem{de_InstrSCI_2010}
de Jong, T. Cognitive load theory, educational research, and instructional design: some food for thought. {\em Instr. Sci.} {\bf 2010}  {\em 38}, 105--134. https://doi.org/10.1007/s11251-009-9110-0
%

\end{thebibliography}
\end{document}